\documentclass[aps,twocolumn,showpacs,preprintnumbers]{revtex4}
\usepackage{CJKutf8}
\usepackage[utf8]{inputenc}
\usepackage[unicode]{hyperref}
\usepackage{graphicx}  % Include figure files
\usepackage{multirow}

\usepackage{fancyhdr}
\usepackage{longtable}
\usepackage{parskip}
\usepackage{dcolumn}   % Align table columns on decimal point

\usepackage{bm}        % bold math
\usepackage{amsfonts}  % Common math fonts
\usepackage{amsmath}   % Common math functions
\usepackage{amssymb}   % Common math symbols

\usepackage{mathrsfs}
\usepackage{mathtools}

\usepackage{slashed}
\usepackage{color}
\usepackage{xcolor}
\usepackage{appendix}

\usepackage{tikz}
\usepackage{mathrsfs}

\usepackage{upgreek}
\usepackage[binary-units=true,per-mode=symbol]{siunitx}
\begin{document}
\title{Measurement of Muon-induced Neutron Yield at the China Jinping Underground Laboratory}% Force line breaks with \\
\thanks{Supported in part by the National Natural Science Foundation of China (11620101004, 11475093), the Key Laboratory of Particle \& Radiation Imaging (TsinghuaUniversity), the CAS Center for Excellence in Particle Physics (CCEPP), and Guangdong Basic and Applied Basic Research Foundation (2019A1515012216). Portion of this work performed at Brookhaven National Laboratory is supported in part by the United States Department of Energy (DE-SC0012704)\\
$ † $ E-mail: orv@tsinghua.edu.cn}%

\author{\begin{CJK}{UTF8}{gbsn} Lin Zhao (赵林)$ ^{1,2} $ Wentai Luo(骆文泰)$ ^{6} $ Lars Bathe-Peters$ ^{1,4} $ Shaomin Chen(陈少敏)$ ^{1,2,3} $ Mourad Chouaki$ ^{1,5} $ Wei Dou(窦威)$ ^{1,2} $ Lei Guo(郭磊)$ ^{1,2} $ Ziyi Guo(郭子溢)$ ^{1,2} $ Ghulam Hussain$ ^{1,2} $ Jinjing Li(李进京)$ ^{1,2} $ Ye Liang(梁晔)$ ^{1,2} $ Qian Liu(刘倩)$ ^{6} $ Guang Luo(罗光)$ ^{7} $ Ming Qi(祁鸣)$ ^{8} $ Wenhui Shao(邵文辉)$ ^{1,2} $ Jian Tang(唐健)$ ^{7} $ Linyan Wan(万林焱)$ ^{1,2} $ Zhe Wang(王\hbox{\scalebox{0.5}[1]{吉}\kern-.1em\scalebox{0.5}[1]{吉}})$ ^{1,2,3} $ Yiyang Wu(武益阳)$ ^{1,2} $ Benda Xu(续本达)$ ^{1,2,3†} $ Tong Xu(徐彤)$ ^{1,2} $ Weiran Xu(徐蔚然)$ ^{1,2} $ Yuzi Yang(杨玉梓)$ ^{1,2} $ Minfang Yeh$ ^{9} $ Aiqiang Zhang(张爱强)$ ^{1,2} $ Bin Zhang(张彬)$ ^{1,2} $ \end{CJK}}

%\author{Lin Zhao$ ^{1,2†} $ Wentai Luo$ ^{6} $ Lars Bathe Peters$ ^{1,4} $ Shaomin Chen$ ^{1,2,3} $ Mourad Chouaki$ ^{1,5} $ Wei Dou$ ^{1,2} $ Lei Guo$ ^{1,2} $ Ziyi Guo$ ^{1,2} $ Ghulam Hussain$ ^{1,2} $ Jinjing Li$ ^{1,2} $ Ye Liang$ ^{1,2} $ Qian Liu$ ^{6} $ Guang Luo$ ^{7} $ Ming Qi$ ^{8} $ Wenhui Shao$ ^{1,2} $ Jian Tang$ ^{7} $ Linyan Wan$ ^{1,2} $ Zhe Wang$ ^{1,2,3} $ Yiyang Wu$ ^{1,2} $ Benda Xu$ ^{1,2,3} $ Tong Xu$ ^{1,2} $ Weiran Xu$ ^{1,2} $ Yuzi Yang$ ^{1,2} $ Minfang Yeh$ ^{9} $ Bin Zhang$ ^{1,2} $ }
\affiliation{%
1. Department of Engineering Physics, Tsinghua University, Beijing 100084, China
}%
\affiliation{%
2. Center for High Energy Physics, Tsinghua University, Beijing 100084, China
}%
\affiliation{%
3. Key Laboratory of Particle \& Radiation Imaging (Tsinghua University), Ministry of Education, China
}%
\affiliation{%
4. Institut für Physik, Technische Universität Berlin, Berlin 10623, Germany
}%
\affiliation{%
5. École Polytechnique Fédérale de Lausanne, Lausanne 1015, Switzerland
}%
\affiliation{%
6. School of Physical Sciences, University of Chinese Academy of Sciences, Beijing 100049, China
}%
\affiliation{%
7. School of Physics, Sun Yat-Sen University, Guangzhou 510275, China
}%
\affiliation{%
8. School of Physics, Nanjing University, Nanjing 210093, China
}%
\affiliation{%
9. Brookhaven National Laboratory, Upton, New York 11973, USA
}%

\collaboration{JNE Collaboration}\noaffiliation

\date{\today}% It is always \today, today,
             %  but any date may be explicitly specified

\begin{abstract}
Solar, terrestrial, and supernova neutrino experiments are subject to muon-induced radioactive backgrounds. The China Jinping Underground Laboratory (CJPL), with its unique advantage of a 2400 m rock coverage and long distance from nuclear power plants, is ideal for MeV-scale neutrino experiments. Using a 1-ton prototype detector of the Jinping Neutrino Experiment (JNE), we detected 343 high-energy cosmic-ray muons and (7.86$ \pm $3.97) muon-induced neutrons from an 820.28-day dataset at the first phase of CJPL (CJPL-I). Based on the muon-induced neutrons, we measured the corresponding muon-induced neutron yield in a liquid scintillator to be $ (3.44 \pm 1.86_{\rm stat.}\pm 0.76_{\rm syst.})\times 10^{-4}\mu ^{-1}\rm g^{-1}cm^{2} $ at an average muon energy of \SI{340}{GeV}. We provided the first study for such neutron background at CJPL. A global fit including this measurement shows a power-law coefficient of (0.75$ \pm $0.02) for the dependence of the neutron yield at the liquid scintillator on muon energy.

\begin{flushleft}
~~~~~~~~~~~~~~~~~~{\bf Keywords:} Underground laboratory, neutrino detector, cosmic-ray muon, neutron yield, liquid scintillator.
\end{flushleft}

\end{abstract}

\pacs{95.85.Ry,~29.40.Mc,~95.55.Vj}
\maketitle

\section{Introduction}
Cosmic-ray muons, produced by the interaction of primary high-energy cosmic rays with the Earth's atmosphere, have a powerful penetrating ability. When such muons pass through mountain rocks and detector materials, they create neutrons. The neutrons are the main source of environmental radiations for underground low-background experiments. They are produced in two ways. The direct production includes muon-nucleus spallation, quasielastic scattering, and negative muon capture by the nucleus. The indirect one is from the muon-initiated electromagnetic and hadronic showers.  The knowledge of cosmic-ray muon-induced neutrons is crucial for the search of rare events associated with neutrino interactions, neutrino-less double-beta decay, and weakly interacting massive particles~(WIMPs) in underground low-background experiments.

In 2000, the CERN Super Proton Synchrotron (SPS) studied the production of the muon-induced neutrons using an artificial muon beam~\cite{hagner2000muon}. The cosmic-ray muon-induced neutrons generated in liquid scintillator (LS) in underground detectors have also been studied by CUBE~\cite{PhysRevC.52.3449}, Daya Bay~\cite{an2018cosmogenic}, Aberdeen Tunnel~\cite{PhysRevD.93.072005}, ASD~\cite{enikeev1987hadrons,malgin2008neutrons,agafonova2013universal}, KamLAND~\cite{abe2010production}, LVD~\cite{lvd-persiani2013measurement}, Borexino~\cite{Borexinobellini2013cosmogenic}, LSD~\cite{aglietta1989neutron}, and many others~\cite{enikeev1987hadrons, bezrukov1973investigation,PhysRevD.62.092005}. A power-law is used to describe the dependence of a neutron yield in an LS on the average muon energy, but measurements up to a few hundreds of GeVs are rare.

Located in Sichuan Province, China, CJPL is one of the deepest underground laboratories in the world~\cite{cheng2017china}. With its unique advantage of a 2400 m rock coverage~\cite{kang2010status} and \SI{\sim 1000}{\kilo m} distance from commercial nuclear power plants, CJPL is ideal for rare-event experiments, such as dark matter searches~\cite{yue2014limits, xiao2014first} and neutrino physics.
%The first phase of CJPL (CJPL-I) opened on 12 December 2010 and the second phase of CJPL (CJPL-II) consists of 4 experimental halls is under construction~\cite{cheng2017china}.
The proposed Jinping Neutrino experiment (JNE)~\cite{beacom2017physics} can serve as an ideal observatory for MeV-scale solar, terrestrial~\cite{PhysRevD.95.053001, wang2020hunting} and supernova~\cite{wei2017discovery} neutrinos.

This paper reports on a measurement of the cosmic-ray muon-induced neutron yield in the LS with an average muon energy of \SI{340}{GeV}~\cite{guo2021muon} using a 1-ton prototype detector of the JNE. The dataset contains 820.28 live days from 31 July 2017 to 27 September 2020. Together with other experiments, this analysis and result contribute to the understanding of the relationship between muon energy and muon-induced neutron yield up to several hundred \si{GeV}.

\section{Prototype Detector and Performance}
The 1-ton prototype detector was constructed in early 2017 to study new neutrino detection techniques and underground backgrounds~\cite{wang2017design}.

\begin{figure}
\includegraphics[width=1\linewidth]{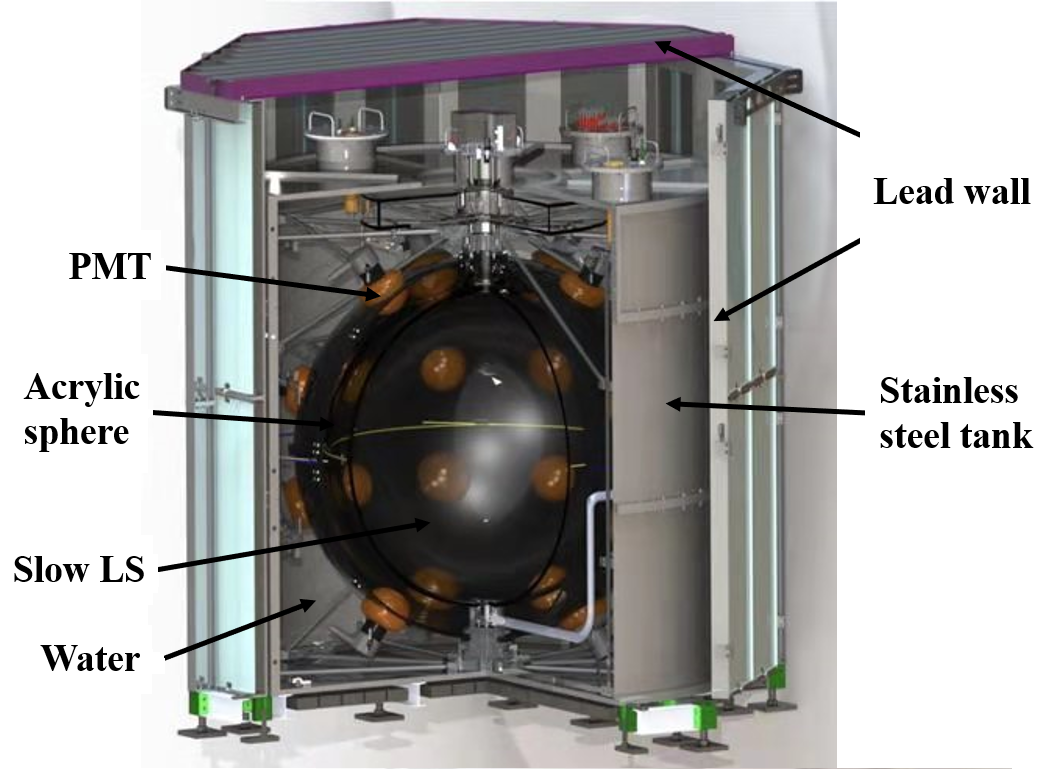}% Here is how to import EPS art
\caption{\label{fig:1tonschema} (color online) Schematic diagram of the 1-ton prototype detector for the Jinping Neutrino Experiment.}
\end{figure}

Figure~\ref{fig:1tonschema} shows a schematic diagram of the prototype detector, which measures \SI{2}{m} in height and diameter. The outermost part is a 5 cm thick lead wall to reduce the environmental background. The innermost part is a \SI{0.645}{m} radius acrylic spherical vessel containing a \SI{1}{\tonne} LS~\cite{li2016separation,guo2019slow}. It consists of linear alkylbenzene (LAB, the molecular formula is $ \rm C_{18}H_{30} $), \SI{0.07}{g\per\litre} of 2,5-diphenyloxazole (PPO), and \SI{13}{mg\per\litre} of the wavelength shifter 1,4-bis(2-methylstyryl)-benzene~(bis-MSB). Thirty 8-inch Hamamatsu R5912 photomultiplier tubes (PMT) outside the acrylic vessel are used to detect light signals.  The PMT gains are maintained at $\sim 10^7$ by adjusting the high voltage power supplies.  A pure water layer between the acrylic sphere and the stainless steel tank serves as passive shielding to suppress the environmental radioactive background.

The electronics system collects and processes the light signal detected by the PMTs. It consists of 4 CAEN V1751 flash ADC boards and one CAEN V1495 logical trigger module. The V1751 board has eight 10-bit ADC channels with \SI{1}{V} dynamic ranges and \SI{1}{GHz} sampling rates. All the PMT signals are digitized in the V1751 boards and then fed into the V1495 boards for logical judgment. We set the PMT-hit discriminator to \SI{10}{mV}, and lowered it to \SI{5}{mV} in phase E after a better understanding of PE charge distribution.  A trigger is issued when more than 25 PMTs are fired simultaneously in a \SI{125}{ns} window. We later lowered the threshold to 10 PMTs in phase G to efficiently detect the events affected by total internal reflections of the acrylic-water boundary. The data acquisition system records \SI{1024}{ns}-sized voltage waveforms of the fired PMTs. For experimental studies, the trigger conditions changed several times. Table \ref{tab:tablePhase} provides the details to define each phase.

We performed a nitrogen purging and sealing experiment on 14 July 2019 to study the laboratory radon leakage into the detector. Meanwhile, we expected that this operation could also remove the oxygen in the LS to reduce the quenching effect, improving the light yield and consequently the energy scale in photo-electrons~(p.e.) per \si{MeV}. Before the nitrogen purging and sealing, the light yield was measured to be 4010 photons per \si{MeV}~\cite{guo2019slow}. This number later increased to 6445.

The 1-ton prototype detector began collecting data on 31 July 2017. Owing to a laboratory maintenance, the entire experiment eventually stopped running on 27 September 2020. We first determined a list of good runs the were neither for pedestal calibration nor detector maintenance. We discarded apparent noise by examining the event rate, single PMT trigger rate, PMT baseline, and baseline fluctuation. The live time after the data quality check was 820.28 days.

\begin{table*}%The best place to locate the table environment is directly after its first reference in text
\caption{\label{tab:tablePhase}%
Running stages and the trigger condition, light yield, and energy scale of each stage of the prototype detector.
}
\begin{ruledtabular}
\begin{tabular}{ccccccccc}

\textrm{DAQ}&
\textrm{Start}&
\textrm{End}&
\textrm{Live time}&
\textrm{Number of}&
\textrm{Window}&
\textrm{Threshold}&
\textrm{Light yield}&
\textrm{Energy scale}\\
phase& & & & PMTs & & & (photon/MeV) & (p.e./MeV) \\
\colrule
A-D&2017.07.31 &2018.10.14 & 392.02 day &25 &1024 ns & 10 mV & 4010 &61.07$ \pm $0.69\\
E-F&2018.10.15 &2019.06.29 & 238.45 day &25 &1024 ns & 5 mV& 4010 &61.07$ \pm $0.69\\
G&2019.06.30 &2019.07.14 & 12.60 day &10 &600  ns & 5 mV& 4010&61.07$ \pm $0.69 \\
H&2019.07.15 &2019.07.22 & 6.61 day &10 &600  ns & 5 mV& 5687&88.20$ \pm $0.49\\
I-J&2019.08.15 &2020.09.27 & 170.60 day &10 &600  ns & 5 mV& 6445 &99.85$ \pm $0.50\\
%I&2019.08.15 &2019.12.11 & 118.17 day &10 &600  ns & 5 mV& 6445 photon/MeV&100.73$ \pm $0.57 (p.e./MeV)\\
%J&2020.09.28 &2020.09.27 & 52.42 day &10 &600  ns & 5 mV& 6445 photon/MeV&97.87$ \pm $0.35 (p.e./MeV)\\
\end{tabular}
\end{ruledtabular}
\end{table*}
%The trigger logic changed to more than 10 PMTs fired with peak larger than 5 mV, and records 600 ns window for higher DAQ efficiency.

\section{Calibration}
We performed several PMT gain and energy scale calibrations for the detector based on the PMT dark noise~(DN) and the decay products of radioactive isotopes in the LS. No dedicated artificial isotopes were deployed because the use of characteristic gamma ray peaks from the contamination to calibrate the small prototype detector was sufficient.

\begin{figure}
  \centering
  {\includegraphics[width=1\linewidth]{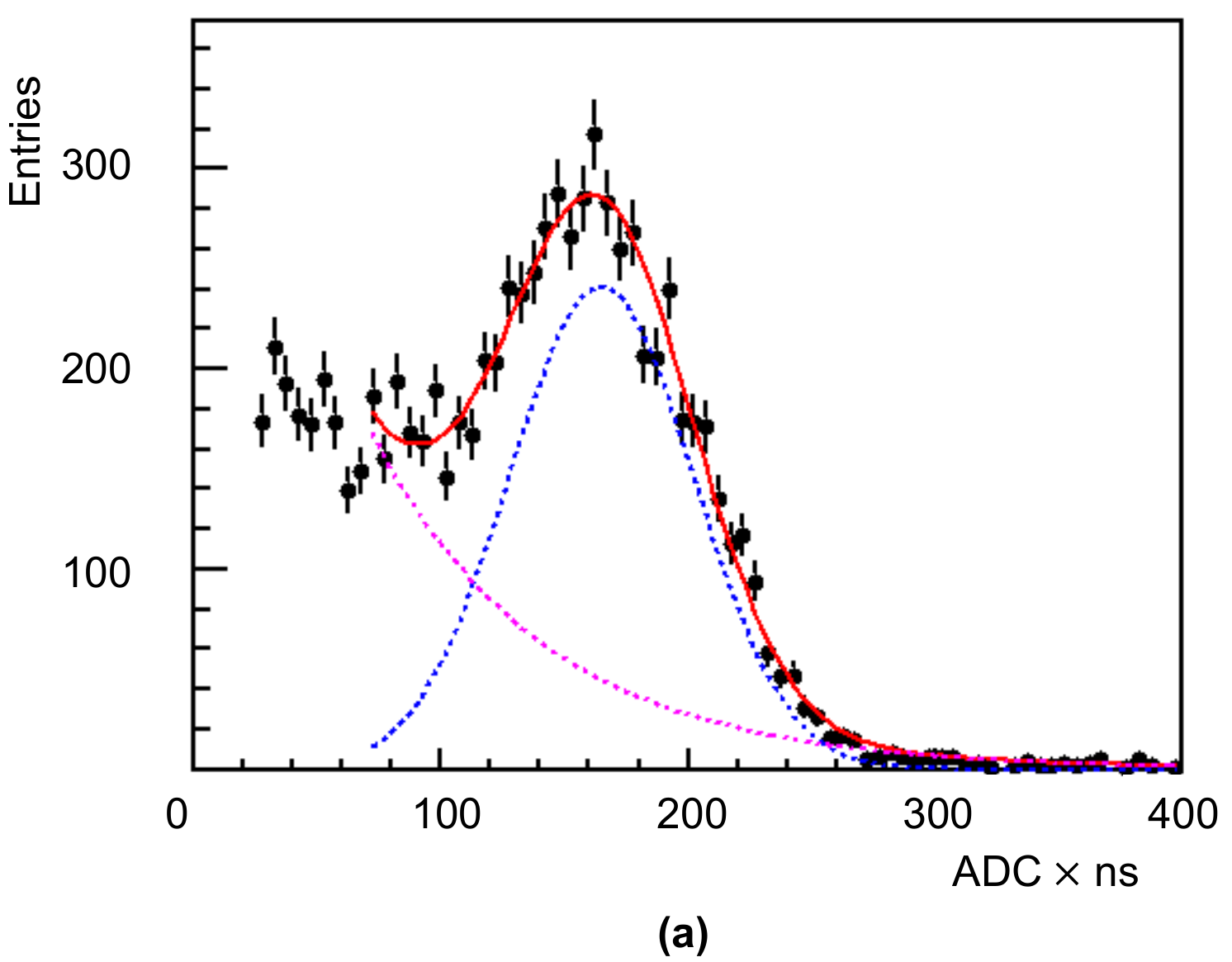}}\\
  {\includegraphics[width=1\linewidth]{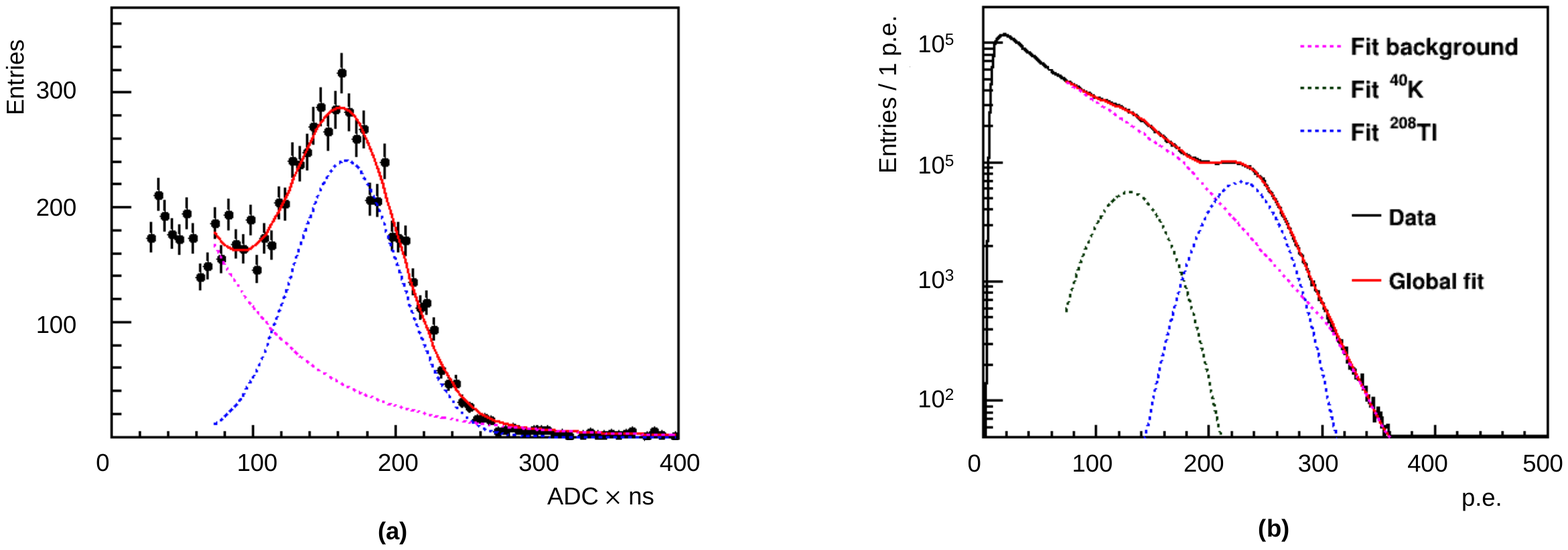}}
  \caption{(color online) Fit example of DN charge spectrum of a PMT (a) and p.e. charge spectrum in one run (b), in which we used three exponential functions to describe the background in three regions.}
  \label{fig:CalibFitResult}
\end{figure}

The thermal electron emission from the PMT photocathode caused the major DN, and they mimicked actual single photoelectron emissions. Since an event time window was 1024 ns wide and the triggered position was between 150 to 600 ns, we selected the DN events from 0 to 150 ns for each PMT. We then fitted the DN charge distribution with a realistic PMT response function~\cite{bellamy1993absolute} to obtain the gain value for each PMT.  Figure~\ref{fig:CalibFitResult}(a) shows an example of the DN charge spectrum.

With the gain value for each PMT of each run, we processed all the data and obtained the total charge for each triggered event. Figure~\ref{fig:CalibFitResult}(b) shows a charge distribution in one run. We extracted the monoenergetic gamma signals emitted by the radioactive isotopes $ ^{40}\rm K $ and $ ^{208}\rm Tl $.

%Figure \ref{fig:esvaration} shows the time variation of the energy scale in terms of p.e./MeV from the $ ^{208}\rm Tl $ decay signal (black dots). The solid red lines denote the average energy scale in each DAQ phase, and the dashed blue line represents the phase border.

By scaling the MC simulations $^{40}\rm K $ and $^{208}\rm Tl$ gammas to the data, we estimated the light yields and energy scales in each DAQ phase Table \ref{tab:tablePhase}. They increased together owing to oxygen removal by nitrogen purging in July 2019. Three sources of energy scale uncertainty were identified: the inconsistency between the $ ^{40}\rm K $ and $ ^{208}\rm Tl $ scales and their fitting errors~(\SI{1.5}{\percent}), background shape uncertainty as shown in Fig.~\ref{fig:CalibFitResult}(b)~(\SI{2.34}{\percent}), and the PMT gain variation with time (\SI{2.84}{\percent}). All three contributions produced a light yield uncertainty of \SI{4}{\percent}.

%\begin{figure}
%\includegraphics[width=1\linewidth]{Fig/EStimevariation.png}% Here is how to import EPS art
%\caption{\label{fig:esvaration} (color online) The time variation of energy scale in terms of p.e./MeV from $ ^{208}\rm Tl $ decay signal (black dots). The solid red lines denote the average energy scale in each DAQ phases, and the dashed blue line denotes the phase border.}
%\end{figure}

\section{Simulation}
\label{Sec:Simulation}
The muon-induced neutron yield refers to the number of neutrons produced by each muon per unit path length and density of the material. Some neutrons produced in the LS escape without being observed, while some detected neutrons are induced by muons traversing other detector components and the rock outside the detector. Considering the above processes, the corrections of the number of muon-induced neutrons in this analysis were estimated by a study based on Geant4 simulation~\cite{agostinelli2003nucl,allison2006geant4}. We determined the average path length of muons through the LS using the Geant4 simulation with the input of muon angular reconstruction result~\cite{guo2021muon}. In the following, we will discuss several important aspects of the simulation.
% The muon path length depends on the muon directional measurement in the detector. The reconstruction to the muon direction is quite challenging at CJPL due to the complicated mountain geometry. The corrections to the number of muon-induced neutrons are necessary to determine the yield.
\subsection{Muon simulation in mountain}
We used the Geant4-based simulation package by Guo~et al.~\cite{guo2021muon} to simulate the muon penetration in the mountain rock to predict various underground muon characteristic profiles. This simulation used Geant4's own standard electromagnetic and muon-nucleus processes. The Jinping mountain terrain data were obtained from the NASA SRTM3 dataset~\cite{farr2007shuttle} with a rock density of \SI{2.8}{g\per cm \cubic}~\cite{wu2013measurement}. The composition of the rock in the simulation utilized the values from the abundance of elements in Earth's crust (percentage by weight)~\cite{lide2004crc}: oxygen (\SI{46.1}{\percent}), silicon(\SI{28.2}{\percent}), aluminum (\SI{8.2}{\percent}), and iron (\SI{5.6}{\percent}). Gaisser's formula~\cite{gaisser2016cosmic,eidelman2004review} reasonably describes the muon flux at sea level. We adopted the modified version from the Daya Bay Collaboration~\cite{guan2006muon} for this simulation, as it has a better flux description for low energy.  It features large zenith angle ranges by parameterizing the kinetic energy $ E $ and zenith angle $ \theta $ of cosmic-ray muons. Figure~\ref{fig:NeutronAngular-compare} shows the simulated angular distribution of underground muons.
	
\subsection{Detector and neutron simulation}
\label{Sec:Detector}
We used Geant4 to generate the incident muon sample and neutrons along the muon's track. The details of the simulated neutrons and their interactions with matter inside the detector are described in this section.

The detector simulation package was a Geant4-based framework. We adopted the same underground muon energy spectrum and angular distribution as~\cite{guo2021muon} to estimate the neutron detection efficiency, and we used a set of empirical formulas from~\cite{wang2001predicting} to generate the neutrons induced by muons. The induced neutron spectrum as a function of the mean energy of muon $ E_{\mu} $ in GeV is given by

\begin{equation}
\label{eq:neutronSpec}
  \frac{\mathrm{d}N}{\mathrm{d}E_{\rm n}}\propto\frac{e^{-7E_{\rm n}}}{E_{\rm n}}+\left(0.52-0.58e^{-0.0099E_{\mu}}\right)e^{-2E_{\rm n}},
\end{equation}
where both $E_\mu$ and $E_\mathrm{n}$ are in \si{GeV}. The neutron zenith angular $\cos\theta$ distribution with respect to the muon is
\begin{equation}
  \frac{\mathrm{d}N}{\mathrm{d}\cos\theta}\propto\frac{1}{(1-\cos\theta)^{0.6}+0.699E_{\mu}^{-0.136}}.
  \label{eq:Neutrontheta_Sim_wang}
\end{equation}	

\begin{figure}
  \centering
  {\includegraphics[width=0.93\linewidth]{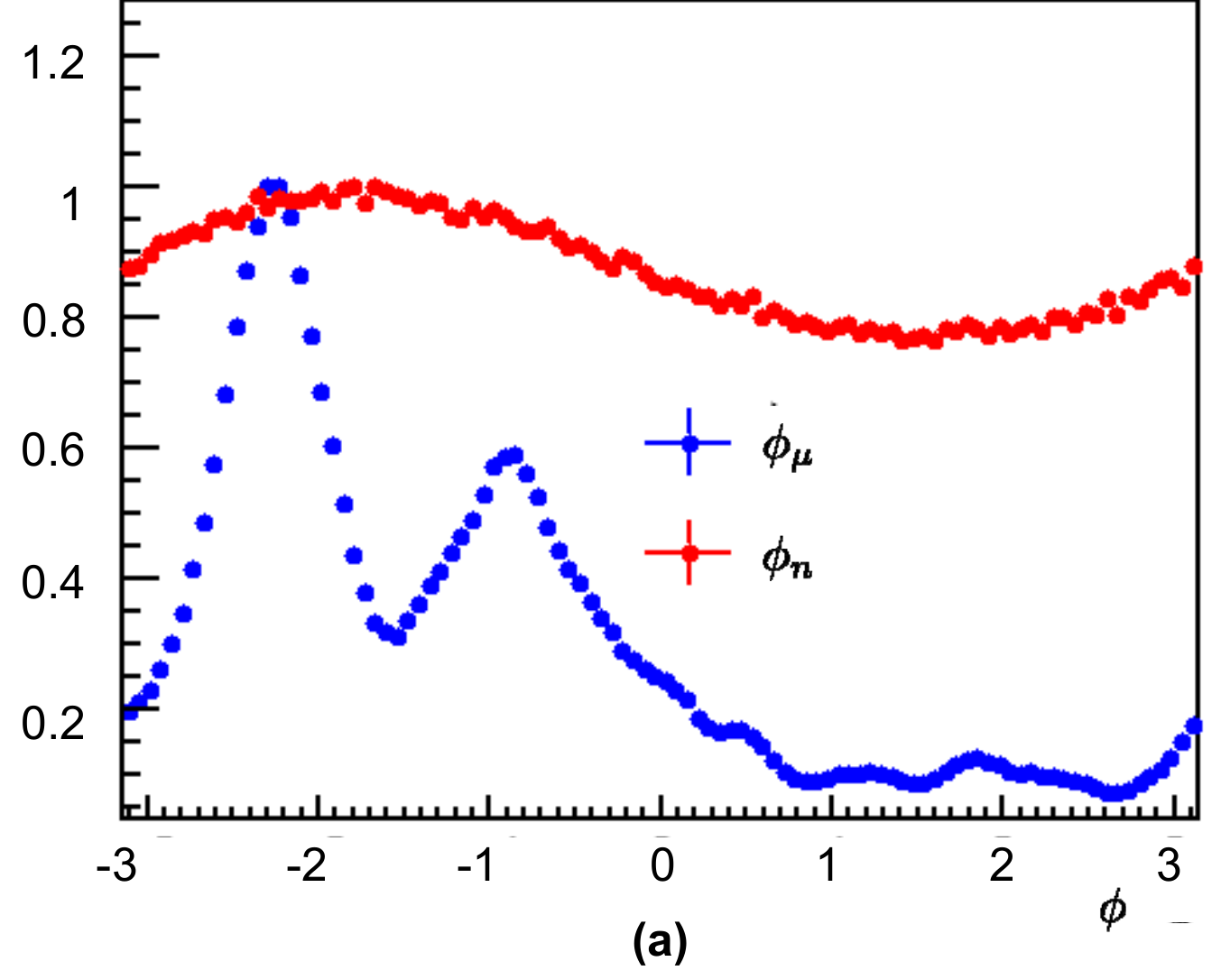}}\\
  {\includegraphics[width=0.93\linewidth]{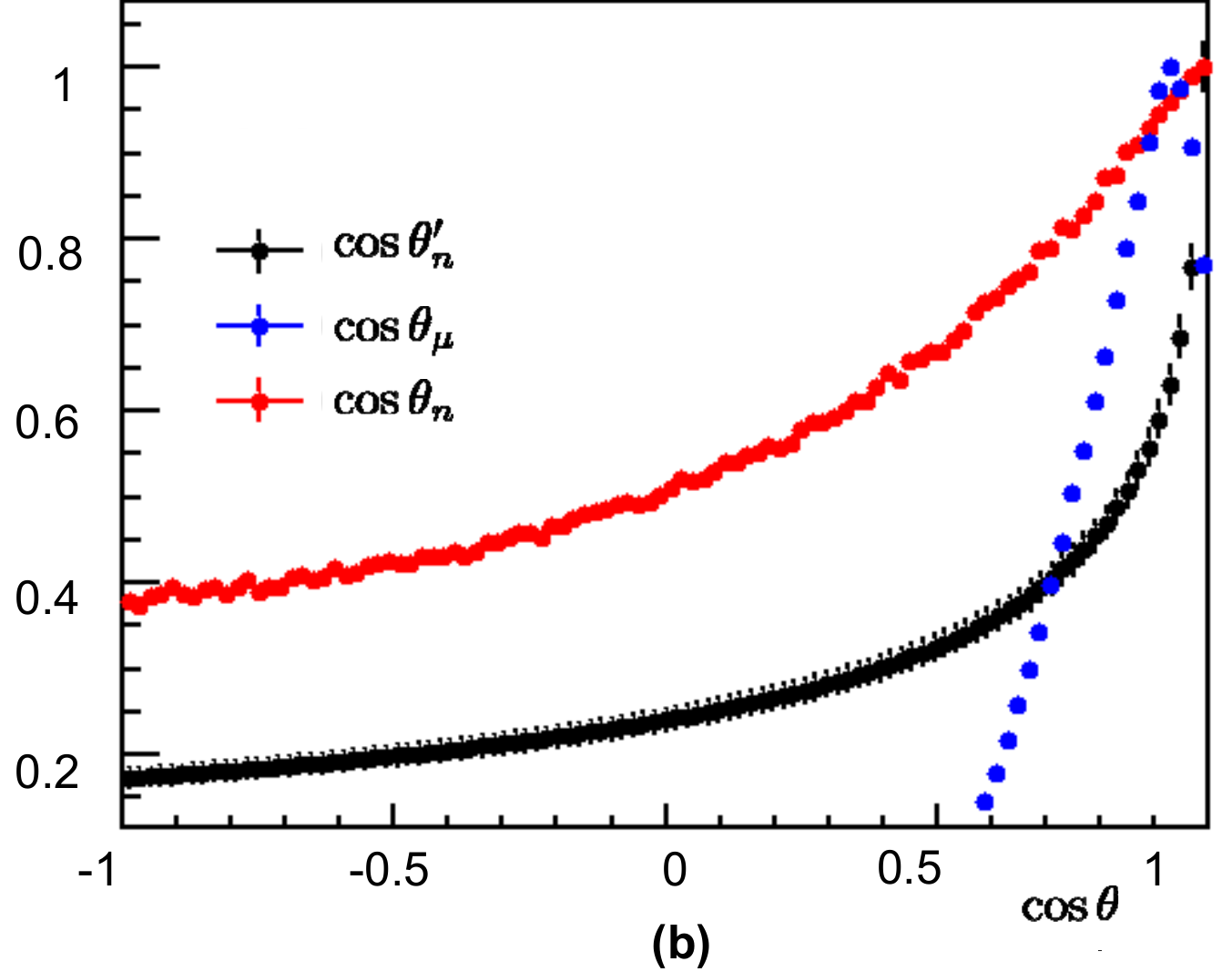}}
  \caption{(color online) Angular distribution of the muon and muon-induced neutron. These two figures show the azimuth angular distributions of the muon ($ \phi_{\mu} $) and neutron ($ \phi_{\rm n} $), the zenith angular distribution of the muon ($ \theta_{\mu} $) and neutron ($ \theta_{\rm n} $), and the angular distribution of the neutron relative to the muon ($ \theta_{\rm n}' $) according to Eq.~\eqref{eq:Neutrontheta_Sim_wang}. All variables are in the laboratory frame.}
  \label{fig:NeutronAngular-compare}
\end{figure}

We assumed that the induced neutron azimuth angular $ \phi $ distribution relative to the parent muon is uniform. Figure~\ref{fig:NeutronAngular-compare} shows the angular distributions of muons and neutrons according to Eq.~\eqref{eq:Neutrontheta_Sim_wang}.

\section{Neutron Yield}
\subsection{Principle}
In this analysis, we selected muon events to search for the associated neutrons. The neutron yield ($ Y_{n} $) in the LS can be expressed as
\begin{equation}
  Y_{n} = \dfrac{N_{n, \rm LS}}{N_{\mu}L\rho},
  \label{eq:yield}
\end{equation}
where $ N_{n, \rm LS} $ is the number of neutrons produced by the muons that traverse the LS target, $ N_{\mu} $ is the number of these muons, $ L $ is the track length of muons in the LS, and $ \rho $ is the density, which was measured to be 0.86 $ \rm g/cm^{3} $.
%Since the chemical difference is tiny between the LS and the slow LS, we do not distinguish them in this analysis.

Out of $ N_{n, \rm LS} $ neutrons, $ N'_{n, \rm LS} $ neutrons are triggered with the efficiency $ \varepsilon_{_{\rm LS}} $:
\begin{equation}
  N_{n, \rm LS}=\frac{N'_{n, \rm LS}}{\varepsilon_{_{\rm LS}}}.
  \label{eq:Efficorrection}
\end{equation}
$ N'_{n, \rm LS} $ includes two contributions: neutrons eventually captured in the LS and those escaping to other detector parts but still producing sufficient photons to become triggered.

Because the detected neutrons could emanate from all the detector components and the rock outside the detector, we define the total number of detected neutrons as,
\begin{equation}
  N'_{n} = \sum_{i} N'_{n, i},
  \label{eq:Qcorrectiondet}
\end{equation}
where $ {i} $ is the index of the materials, including the LS, water, iron tank, lead, and rock. $ N'_{n, \rm LS} $ is given by:
\begin{equation}
  N'_{n, \rm LS}=\xi_{_{\rm LS}} N'_{n},
  \label{eq:Qcorrection}
\end{equation}
where $ \xi_{_{\rm LS}} $ is the fraction of detected neutrons generated in the LS.

According to Eqs.~\eqref{eq:yield}, \eqref{eq:Efficorrection} and \eqref{eq:Qcorrection}, the muon-induced neutron yield is finally expressed as
\begin{equation}
  Y_{n} = \dfrac{\xi_{_{\rm LS}} N'_{n}/\varepsilon_{_{\rm LS}}}{N_{\mu}L\rho}
  \label{eq:yieldFinal}
\end{equation}

The following sections present the selection criteria for the muon-induced neutrons and the determination of all the quantities shown in Eq.~\eqref{eq:yieldFinal}.

\subsection{Muon event selection and flux}
\begin{figure}
  \centering
  {\includegraphics[width=1\linewidth]{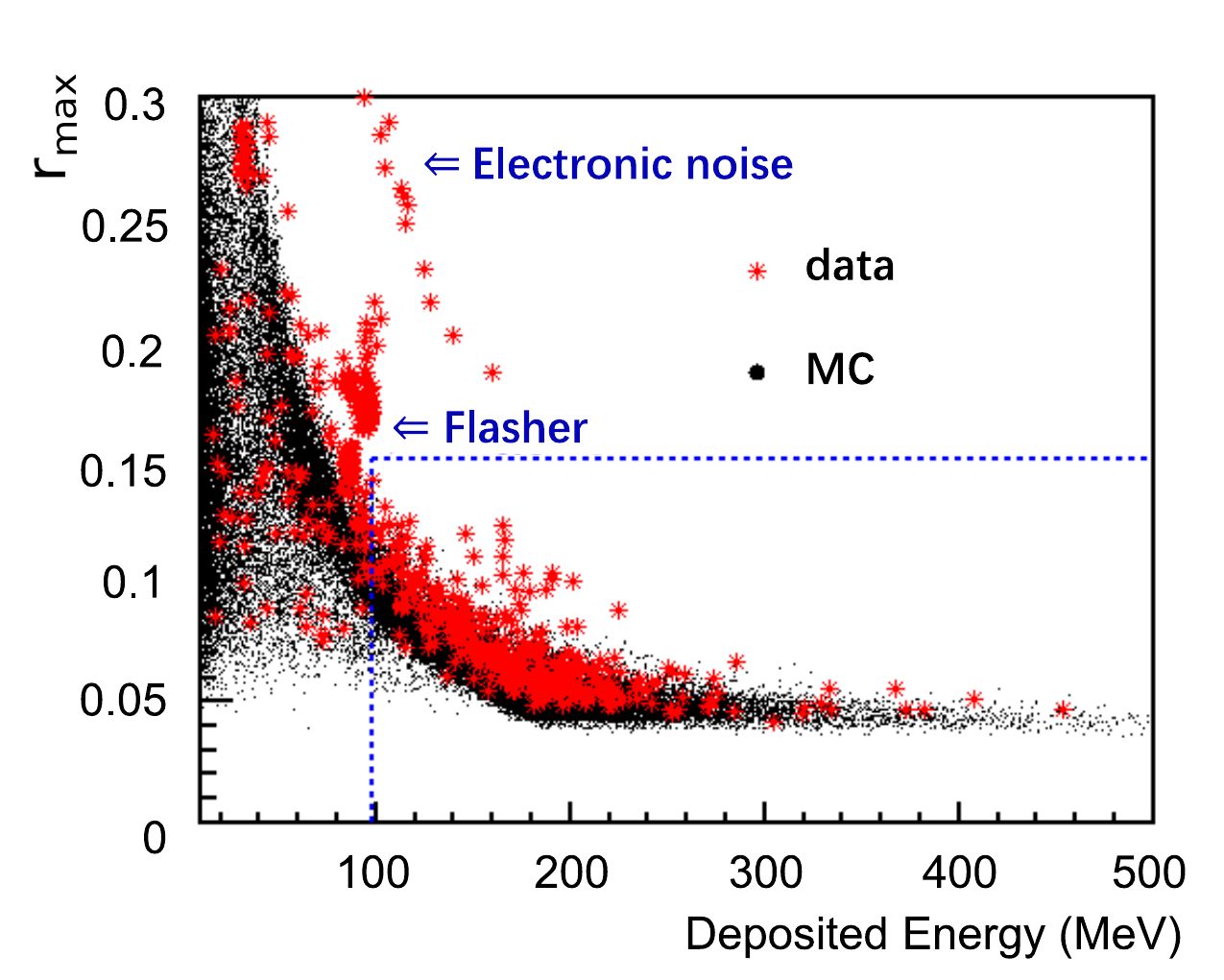}}
  \caption{(color online) Scatter plot of deposited energy and $ r_{\rm max} $; the black dots represent the muon events from the MC simulation, and the red dots represent the muon candidates selected from the experimental data. Three hundred forty-three selected muon events spread in the region of $ E_{\rm dep}>\SI{98}{MeV}$ and $ r_{\rm max}<0.15 $, while flasher and electronic noise events had larger $ r_{\rm max} $ and relatively smaller $ E_{\rm dep} $ values. We also removed low energy events that may contain indistinguishable radioactive background, shower, and noise events.}
  \label{fig:MuonE-R}
\end{figure}

We required a minimum energy deposit of \SI{98}{MeV} for each muon event. We cut out the electronics noise and flasher events by requiring the ratio of the maximum p.e.~among the PMTs to the total p.e.~in each event, denoted by $ r_{\rm max} $, to be less than 0.15. Figure~\ref{fig:MuonE-R} shows a two-dimensional distribution of the deposited energy $ E_{\rm dep} $ and $ r_{\rm max} $. There were 343 high-energy cosmic-ray muons, i.e., $ N_{\mu} $ in Eq.~\eqref{eq:yieldFinal}. Using the same method as Ref.~\cite{guo2021muon} and an enlarged exposure, we updated the muon flux of CJPL-I to $ \rm (3.61\pm 0.19_{stat.}\pm 0.10_{sys.})\times 10^{-10}cm^{-2}s^{-1} $.

%Together with the MC simulation, the estimated muon flux and average muon energy of CJPL-II are shown in Table \ref{tab:CJPL2muonpar}.

%A template-based method to construct the muon direction were used in muon flux research\cite{guo2021muon}. Figure Updated muon candidates together with improved time calibration were used

\subsection{Neutron event selection}
	We selected events for the neutron capture on hydrogen following a tagged parent muon. The neutron capture time $ t $ distribution follows
\begin{equation}
  P(t)\propto e^{-t/\tau_{n}},
  \label{eq:NeutronCapTime}
\end{equation}	
where $ \tau_{n}=\SI{216}{\micro s}$ is the mean neutron capture time. It was calculated using the LS's elemental composition and the thermal neutron capture cross-sections from~\cite{mughabghab2012neutron} at the detector temperature.

\begin{figure}
  \centering
  {\includegraphics[width=1\linewidth]{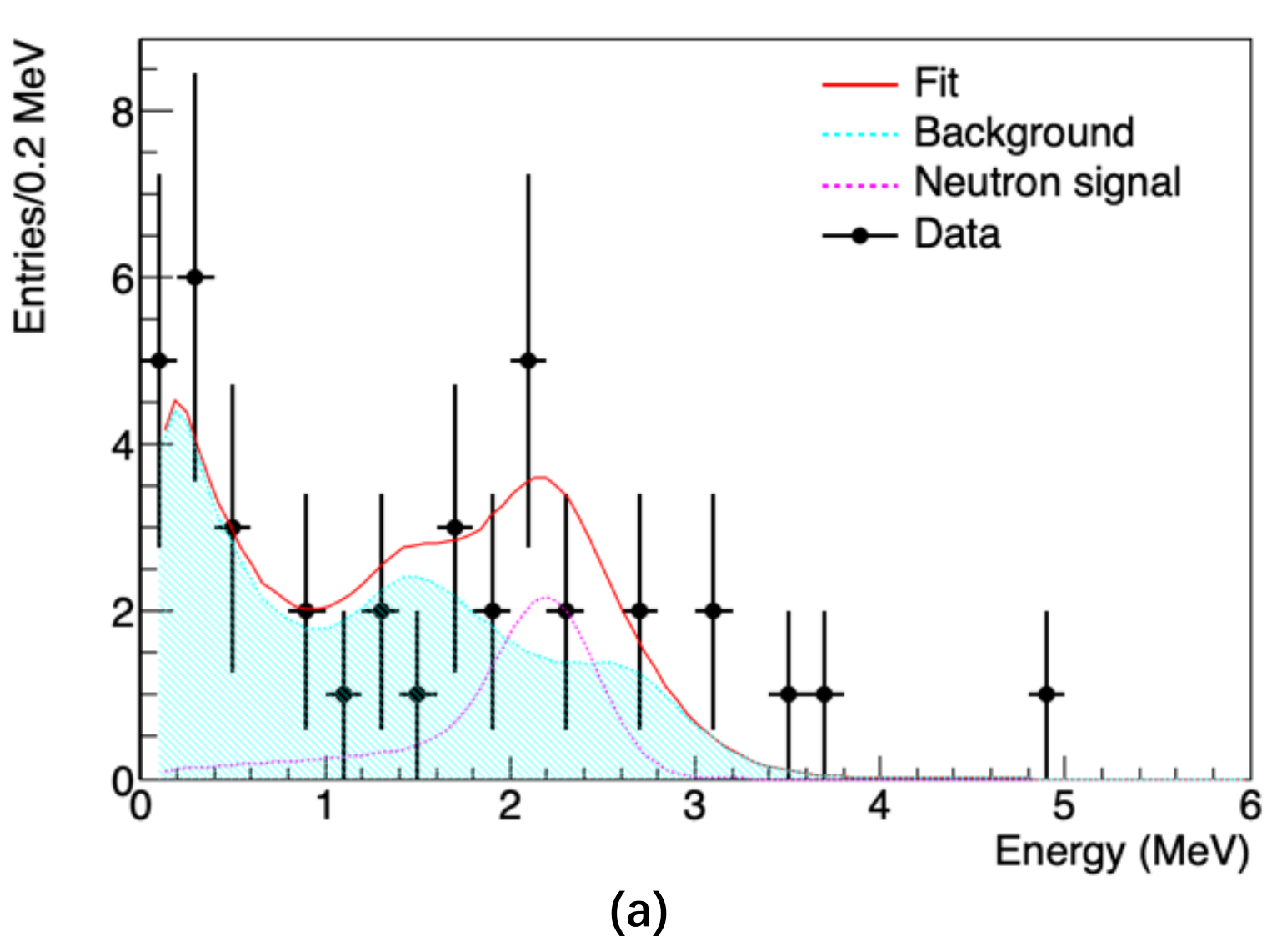}}\\
  {\includegraphics[width=1\linewidth]{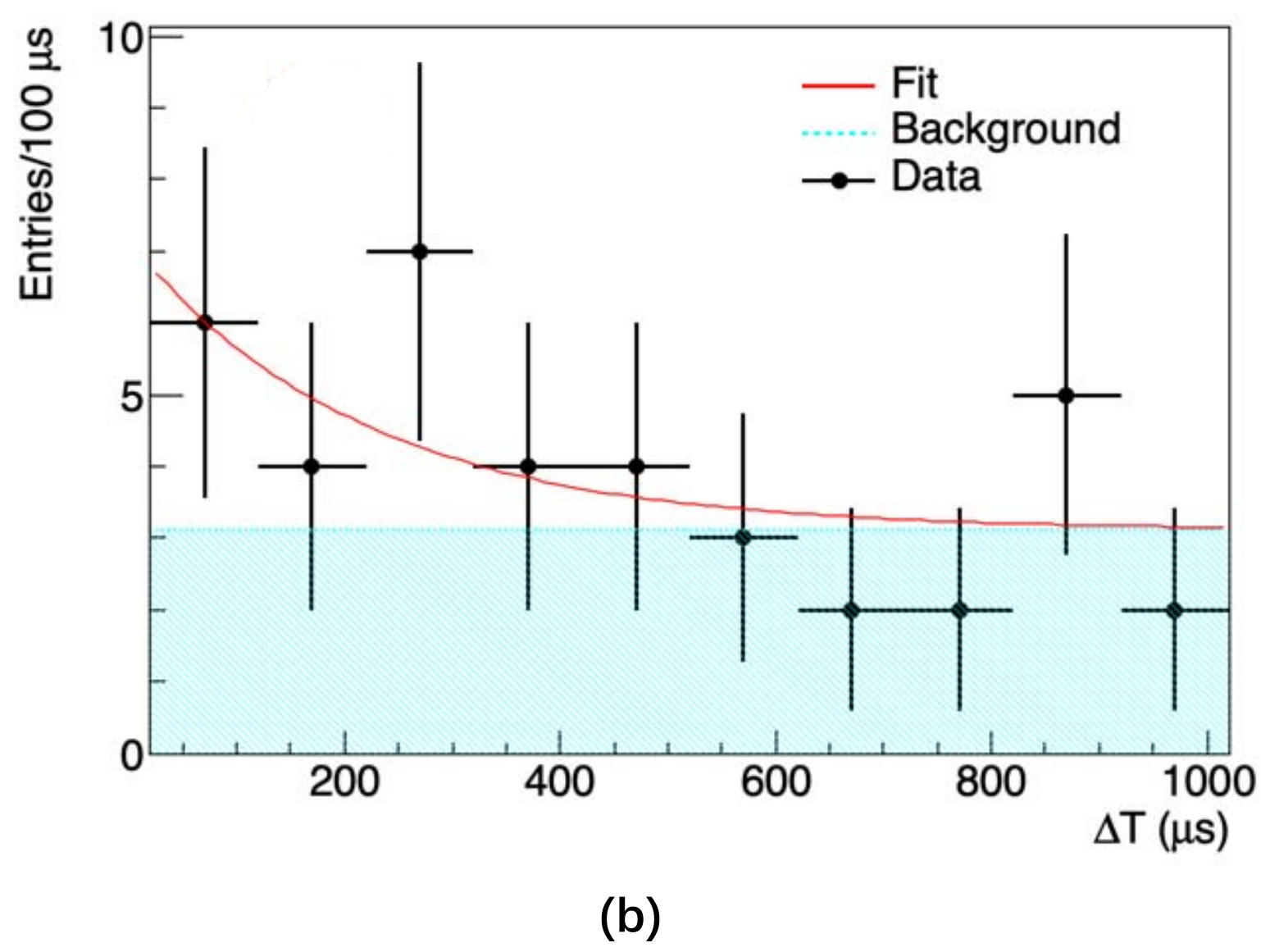}}
  \caption{(color online) Fit of energy (a) and time (b) spectra of neutron capture. The data points were captured from events in the $\SI{20}{\mu s}<\Delta {T}<\SI{1020}{\mu s}$ window after the muon.  Background distributions are scaled from time sideband data. The neutron signal energy spectrum was estimated using Monte Carlo detector simulation. The dataset was a simple addition of the low and high threshold periods.}
  \label{fig:FitResult}
\end{figure}

To avoid electronics-induced baseline distortions, we define the signal window to start from \SI{20}{\micro s} after the muon's passage and end at \SI{1020}{\micro s} to cover neutron capture time, shown in Fig.~\ref{fig:FitResult} as ``data''.  The ``backgroud'' in Fig.~\ref{fig:FitResult} is scaled from the $ \SI{1020}{\micro s}<\Delta {T}<\SI{5001020}{\micro s}$ time sideband window following the parent muons.  A \SI{2.2}{MeV} peak was evident in the energy spectrum, but it did not occur in the background. Three peaks were observed in the background energy spectrum. The two at approximately 0 MeV and 1.5 MeV were due to the trigger condition changes. In phases A-F, the 25-PMT threshold was just below \SI{1.5}{MeV}. The one at approximately 2.6 MeV was owing to the radioactive $ \rm ^{208}Tl $ background. We evaluated the number of muon-induced neutrons $ N'_{n} $ with an unbinned maximum likelihood fit incorporating both the neutron-like event energy and time to the parent muon in this analysis.

%Fig.~\ref{fig:FitResult} (a) shows the energy spectrum in signal window $ 20 \ \upmu {\rm s}<\Delta {T}<1020\  \upmu {\rm s} $ (black crosses) and energy spectrum in the background window $ 1020 \ \upmu {\rm s}<\Delta {T}<5001020 \ \upmu {\rm s} $ which have been normalized to 1000 $ \upmu \rm s $ (dash blue line) following muon events. The energy spectrum shows a peak around 2.2 MeV, which is not evident in the expected background distribution.

%\begin{figure}
%  \centering
%   {\includegraphics[width=0.8\linewidth]{fig/E3sigma.png}}
%  \caption{(color online) Energy spectrum of the events following muons in the signal time window $ 20 \ \upmu {\rm s}<\Delta {T}<1020 \ \upmu {\rm s} $ (black crosses) and normalized energy spectrum in background window $ 1020 \ \upmu {\rm s}<\Delta {T}<501020 \ \upmu {\rm s} $ (solid blue line). The energy spectra clearly shows a peak from neutron captures on $ \rm ^{1}H $ (2.225 MeV), which are not evident in the background.}
%  \label{fig:E3sigma}
%\end{figure}

We describe the energy spectrum using the following probability density function:
\begin{equation}
  f(E)=(1-\omega)f_{\rm sig}+\omega f_{\rm bkg},
  \label{eq:Efunc}
\end{equation}
where $ \omega $ is the ratio of the background rate to the total event rate in the signal time window $ 20 \ \upmu {\rm s}<\Delta {T}<1020 \ \upmu {\rm s} $, $ f_{\rm bkg} $ is the background probability, and $ f_{\rm sig} $ is the signal probability~\cite{cheng2016determination}:
%\begin{widetext}
\begin{equation}
  f_{\rm sig}=\alpha\frac{1}{\sigma\sqrt{2\pi}}e^{-\frac{(E-\mu)^{2}}{2\sigma^{2}}}+(1-\alpha)\frac{\lambda e^{\frac{\sigma^{2}\lambda^{2}+2\lambda E}{2}}}{e^{\lambda\mu}-1}f_{\rm erf},
  \label{eq:Calfunc}
\end{equation}
%\end{widetext}
in which
\begin{equation}
  f_{\rm erf}={\rm erf}\left(\frac{\mu-E-\sigma^{2}\lambda}{\sqrt{2}\sigma}\right)-{\rm erf}\left(\frac{-E-\sigma^{2}\lambda}{\sqrt{2}\sigma}\right),
  \label{eq:Calfuncerf}
\end{equation}
where $ \mu $ and $ \sigma $ are the mean and standard deviation of the expected 2.2 MeV neutron-hydrogen capture peak, respectively, $ \alpha $ is the signal fraction, $ \lambda $ is the slope of the exponential tail caused by energy leak and edge effect, and $\rm erf $ is the Gaussian error function: $ {\rm erf}(E)=\frac{2}{\sqrt{\pi}}\int_{0}^{E} e^{-t^{2}}dt $. Equation~\eqref{eq:Calfunc} describes the MC energy spectrum very well, and the parameters were determined from the MC energy spectrum. Since the detector ran in several different trigger conditions, the final MC energy spectrum was the average of those weighted by the live time for each DAQ phase.

%By comparing the event number of MC energy spectrum and integrated number of fitting result from equation \eqref{eq:Calfunc}, the systematic uncertainty of neutron number are estimated to be 1\%.
%Figure \ref{fig:MCEnergyFit} shows the fit result.
%\begin{figure}
%  \centering
%   {\includegraphics[width=1\linewidth]{fig/nHSimEnergy}}
%  \caption{(color online) Fitted result of MC nH energy spectrum by equation \eqref{eq:Calfunc}. The black point is the average energy spectrum weighted by the live time for each DAQ phase, the right line is the fit result.}
%  \label{fig:MCEnergyFit}
%\end{figure}

The time spectrum of neutron capture is described by
\begin{equation}
  N(t)=\frac{N'_{n}}{\tau_{n}}e^{-(t-t_{\mu})/\tau_{n}}+N_{\rm bkg},
  \label{eq:Tfunc}
\end{equation}
where $ N'_{n} $ is the total number of neutron captures associated with the selected signal window, and $ N_{\rm bkg} $ is the background rate. Considering the selection of signal time window, we have the probability density function
\begin{equation}
  f(t)=(1-\omega)f_{\rm sig}^{'}+\omega f_{\rm bkg}^{'},
  \label{eq:TfuncPdf}
\end{equation}
where,
\begin{equation}
  f_{\rm sig}^{'}=\frac{\frac{1}{\tau_{n}}e^{-(t-t_{\mu})/\tau_{n}}}{\int_{20\ \upmu \rm s}^{1020\ \upmu \rm s}\frac{1}{\tau_{n}}e^{-(t-t_{\mu})/\tau_{n}}{\rm d}t}, \\
\end{equation}
and
\begin{equation}
  f_{\rm bkg}^{'}=\frac{1}{\int_{\SI{20}{\micro s}}^{\SI{1020}{\micro s}}{\rm d}t}.
  \label{eq:Tfunc-bkg-sig}
\end{equation}
which are the normalized signal and background time spectra, respectively.

We consider that the number of neutron-capture events follows a Poisson distribution with the expected value $ \nu $. Since the energy and time of neutron captures are independent, the 2-dimensional likelihood function is
\begin{equation}
  \mathcal{L}=\frac{\nu^{m}}{m!}e^{-\nu}\prod \limits_{i=1}^{m}\left[f(E_{i})f(t_{i})\right].
  \label{eq:Likefunc}
\end{equation}
where $ m = 39 $ is the number of neutron candidates in the signal window.

%The $ \varepsilon_{_{\rm LS}} $ and $ \xi_{_{\rm LS}} $ in Eq~\eqref{eq:yieldFinal} are affected by trigger conditions.

We divided the DAQ phases into two parts: the high threshold period and the low one, namely DAQ A-F and DAQ G-J. Analyzed separately, the numbers of muon-induced neutrons were observed to be $ 6.37
\pm3.46_{\rm stat.} $ and $ 1.49\pm1.93_{\rm stat.} $, respectively. The total number of muon-induced neutrons $ N'_{n} $ was calculated as $ 7.86\pm3.97_{\rm stat.} $, which shows the observed neutrons with a statistical significance at the 2$ \sigma $ level. Figure~\ref{fig:FitResult} shows the fit of energy (a) and time (b) spectra of neutron capture against the combined dataset of the two periods.
%In order to study the significance of the neutron signal, we fitted the neutron candidates using pure background 2-dimensional likelihood function, i.e., Eq.~\eqref{eq:Likefunc} with fixed $ \omega = 1 $ in Eq.~\eqref{eq:Efunc} and \eqref{eq:TfuncPdf}. This scenario gave the maximum likelihood result $ -2{\rm log}\mathcal{L}_{\rm max}(b) = 445.12 $. Compared with $ -2{\rm log}\mathcal{L}_{\rm max}(s+b) = 441.71 $, the significance of neutron signal is $ 1.8 \ \sigma $.

%In order to study the significance of the neutron signal, we further separate background dominant and signal enhanced samples. The signal enhanced samples are in energy region 1 MeV to 2.8 MeV, and background dominant are in energy region below 1 MeV or above 2.8 MeV. The number of asymmetry events of lifetime distribution are also observed below 500 $ \upmu \rm s $ and above. The number of samples in background dominant region is 12:9, $ (12-9)/\sqrt{21}=0.7 \sigma $; and in signal enhanced region is 13:5, $ (13-5)/\sqrt{18}=1.9 \sigma $. Therefore, the signal is still somewhat evident. All this statistic uncertainty was already reflected in the result.

We estimated the systematic effect on the number of neutrons due to the uncertainties of the parameters obtained from MC simulation and the energy scale to be \SI{0.64}{\percent} and \SI{0.81}{\percent}. They provided a combined systematic uncertainty on $ N'_{n} $ of \SI{1.0}{\percent}.

\subsection{$ \varepsilon_{_{\rm LS}} $, $ L $, and $ \xi_{_{\rm LS}} $}

%The probability of neutrons escaping from the LS region is $ (14.68 \pm 0.65)\% $.

%These two parts were treated separately in the neutron detection efficiency $ \varepsilon_{_{\rm LS}} $ in equation \eqref{eq:yieldFinal}:

%\begin{equation}
%  \varepsilon_{_{\rm LS}}=\left(1-f_{\rm esp}\right)\varepsilon_{_{\rm LS,det}}+f_{\rm esp}\varepsilon_{_{\rm Others,det}}.
%  \label{eq:Efficorrection2parts}
%\end{equation}
%where, $ f_{\rm esp} $ is the probability of neutrons escaping from the LS region, $ \varepsilon_{_{\rm LS,det}} $ is the detection efficiency of neutrons generated and captured in the LS, and $ \varepsilon_{_{\rm Others,det}} $ is the detection efficiency in other areas of the detector with the same cuts. $ f_{\rm esp} = (14.68 \pm 0.65)\% $ keeps consistent within the margin of error.

%thus $ \left(1-f_{\rm esp}\right) $ is the probability of neutrons captured in the LS region,

The MC simulation determined the neutron detection efficiency $ \varepsilon_{_{\rm LS}} $ in Eq.~\eqref{eq:yieldFinal} for each DAQ phase. Owing to the small volume of the LS region in this detector and the relatively large mean free path of muon-induced neutrons, $ \varepsilon_{_{\rm LS}} $ is dependent on the neutrons' distribution positions. We compared the neutrons produced along their parent muons track and those uniformly generated in the LS region and evaluated a \SI{6.65}{\percent} contribution to the systematic uncertainty of $ \varepsilon_{_{\rm LS}} $. Table \ref{tab:tableeffi} shows the result for each DAQ phase. We obtained the weighted average by the live time for $ \varepsilon_{_{\rm LS}} $ used in Eq.~\eqref{eq:yieldFinal}.

%The precision of $ \varepsilon $ is subject to statistics of the experimental data and the complicated structure of the detector.

The underground muon energy and angular spectra and muon selection criteria determined the distribution of $ L $ and its uncertainty. The average track length $ L $ of a muon passing the LS was estimated to be $ 108.01\pm 14.10 \rm \ cm $ from the MC simulation.

%Fig.~\ref{fig:FitResult} shows the track length distribution of muon passing through the whole detector (black), liquid scintillator (red), and the water (blue).

We calculated $ \xi_{_{\rm LS}} $ by simulating the detected number of muon-induced neutrons in each component $ N'_{n,i} $ defined by Eqs.~\eqref{eq:yield}, \eqref{eq:Efficorrection} and \eqref{eq:Qcorrectiondet}. Table \ref{tab:tableeffi} shows the $ \xi_{_{\rm LS}} $ value in each DAQ phase. We obtained the weighted average of $ \xi_{_{\rm LS}} $ for use in Eq.~\eqref{eq:yieldFinal} using the live time.

As shown in Table \ref{tab:tableeffi}, $ \varepsilon_{_{\rm LS}} $ and $ \xi_{_{\rm LS}} $ varied with the trigger conditions and light yields of the detector.

Similar to $ \varepsilon_{_{\rm LS}} $ and $ \xi_{_{\rm LS}} $, we defined the detection efficiency $ \varepsilon_{_{\rm Other}} $ and fraction of detected neutrons $ \xi_{_{\rm Other}} $ generated in other components, which include the water, acrylic vessel, iron tank, lead wall of the detector, and rock in the Jinping Mountain. They all can contribute to the detected neutrons. $ \varepsilon_{_{\rm Other}} $ and $ \xi_{_{\rm Other}} $ are used to understand the contribution of detected neutrons from other components and the detector performance of neutron detection. Table \ref{tab:tableeffix} lists the $ \varepsilon_{_{\rm Other}} $ and $ \xi_{_{\rm Other}} $ values for DAQ phase A-D, as an example. The sum of all the $ \xi $ values equals \SI{100}{\percent}.

\begin{table}%The best place to locate the table environment is directly after its first reference in text
\caption{\label{tab:tableeffi}%
Detection efficiency $ \varepsilon_{_{\rm LS}} $ and fraction factor $ \xi_{_{\rm LS}} $ values in Eq.~\eqref{eq:yieldFinal} for different DAQ phases.
}
\begin{ruledtabular}
\begin{tabular}{ccc}
\textrm{DAQ phase}&
\textrm{$ \varepsilon_{_{\rm LS}} $ (\%)}&
\textrm{$ \xi_{_{\rm LS}} $ (\%)}\\
\colrule
A-D&40.65$ \pm $2.70	&74.01$ \pm $12.53\\
E-F&40.74$ \pm $2.71	&72.67$ \pm $12.24\\
G&53.54$ \pm $3.56	&67.86$ \pm $11.25\\
H&55.17$ \pm $3.67	&61.79$ \pm $10.09\\
I, J&56.42$ \pm $3.75	&61.36$ \pm $9.86\\
\end{tabular}
\end{ruledtabular}
\end{table}

\begin{table*}%The best place to locate the table environment is directly after its first reference in text
\caption{\label{tab:tableeffix}%
Efficiency $ \varepsilon_{_{\rm Other}} $ and fraction factor $ \xi_{_{\rm Other}} $ value of all components for DAQ phases A-D.
}
\begin{ruledtabular}
\begin{tabular}{cccccc}
\textrm{}&
\textrm{Water}&
\textrm{Acrylic}&
\textrm{Iron}&
\textrm{Lead}&
\textrm{Rock}\\
\colrule
$ \varepsilon_{_{\rm Other}} $ (\%)&$ 4.89\pm 0.33 $&$ 18.79\pm 1.25 $&$ 0.58\pm 0.04 $&$ 0.50\pm 0.03 $&$ 0.073\pm 0.005 $\\
$ \xi_{_{\rm Other}} $ (\%)&$ 9.55\pm 2.89 $&$ 1.00\pm 0.27 $&$ 2.02\pm 0.47 $&$ 13.16\pm 2.17 $&$ 0.26\pm 0.07 $\\
\end{tabular}
\end{ruledtabular}
\end{table*}

\subsection{Yield Result}
Table \ref{tab:resultpara} lists all the variables and uncertainties used in Eq.~\eqref{eq:yieldFinal}.  We calculated the neutron yield in the LS of DAQ phases A-F to be $ Y_{n}=(4.84 \pm 2.63_{\rm stat.}\pm 1.08_{\rm syst.})\times 10^{-4}\mu ^{-1}\rm g^{-1}cm^{2} $, and DAQ phases G-J to be $ Y_{n}=(2.03 \pm 2.63_{\rm stat.}\pm 0.44_{\rm syst.})\times 10^{-4}\mu ^{-1}\rm g^{-1}cm^{2} $. These two measurements were statistically independent, but their systematical uncertainties were fully correlated. The combined neutron yield is $ Y_{n}=(3.44 \pm 1.86_{\rm stat.}\pm 0.76_{\rm syst.})\times 10^{-4}\mu ^{-1}\rm g^{-1}cm^{2} $. Owing to maintenance of the laboratory, the experiment has been shut down since September 2020, preventing an opportunity to accumulate more data to improve the statistical precision as desired. A corresponding upper limit of $7.78\times 10^{-4}\mu ^{-1}\rm g^{-1}cm^{2} $ at 99\% C.L. was also indicated.

\begin{table}%The best place to locate the table environment is directly after its first reference in text
\caption{\label{tab:resultpara}%
Parameters used in Eq.~\eqref{eq:yieldFinal} to calculate the neutron yield in the LS.
}
\begin{ruledtabular}
\begin{tabular}{ccccc}
\textrm{DAQ}&
\textrm{Parameter}&
\textrm{Value}&
\textrm{Statistical}&
\textrm{Systematic}\\
\textrm{phase} & & & uncertainty & uncertainty \\
\colrule
\multirow{6}{*}{A-F}
&$  N'_{n} $	&6.37	&3.46	&0.06\\
&$  N_{\mu} $			&256	&-		&-\\
&$  \varepsilon_{_{\rm LS}}  \ (\%) $		&40.68	&-		&2.70\\
&$  \xi_{_{\rm LS}} \ (\%) $				&73.50	&-		&12.42\\
&$  \rho \ (\rm g/cm^{3}) $	&0.86	&-		&-\\
&$  L \ (\rm cm) $				&108.01	&-		&14.10\\
\colrule
\multirow{6}{*}{G-J}
&$  N'_{n} $	&1.49	&1.93 	&0.01\\
&$  N_{\mu} $			&87	&-		&-\\
&$  \varepsilon_{_{\rm LS}}  \ (\%) $		&56.19	&-		&3.73 \\
&$  \xi_{_{\rm LS}} \ (\%) $				&61.81	&-		&9.96\\
&$  \rho \ (\rm g/cm^{3}) $	&0.86	&-		&-\\
&$  L \ (\rm cm) $				&108.01	&-		&14.10\\
\end{tabular}
\end{ruledtabular}
\end{table}

%\begin{table*}%The best place to locate the table environment is directly after its first reference in text
%\caption{\label{tab:CJPL2muonpar}%
%Estimated muon flux, average muon energy, and neutron yield in CJPL-II
%}
%\begin{ruledtabular}
%\begin{tabular}{ccccc}
%\textrm{Parameter}&
%\textrm{CJPL-II}&
%\textrm{CJPL-II}&
%\textrm{CJPL-II}&
%\textrm{CJPL-II}\\
%\textrm{}&
%\textrm{hall A}&
%\textrm{hall B}&
%\textrm{hall C}&
%\textrm{hall D}\\
%\colrule
% Average muon energy (GeV)&327	&322&361 &357\\
%      Muon flux ($ \rm \times 10^{-10}cm^{-2}s^{-1} $)&$  \sim $4.0 &$  \sim $3.8 &$  \sim $2.3	&$  \sim $2.5\\
%      Neutron yield ($ \times 10^{-4}\ \upmu^{-1}\rm g^{-1}cm^{2}  $) &$  \sim $3.5 &$  \sim $3.4 &$  \sim $3.7	&$  \sim $3.7\\
%\end{tabular}
%\end{ruledtabular}
%\end{table*}

\section{Comparison with other experiments}
\begin{figure*}
  \centering
   {\includegraphics[width=0.8\linewidth]{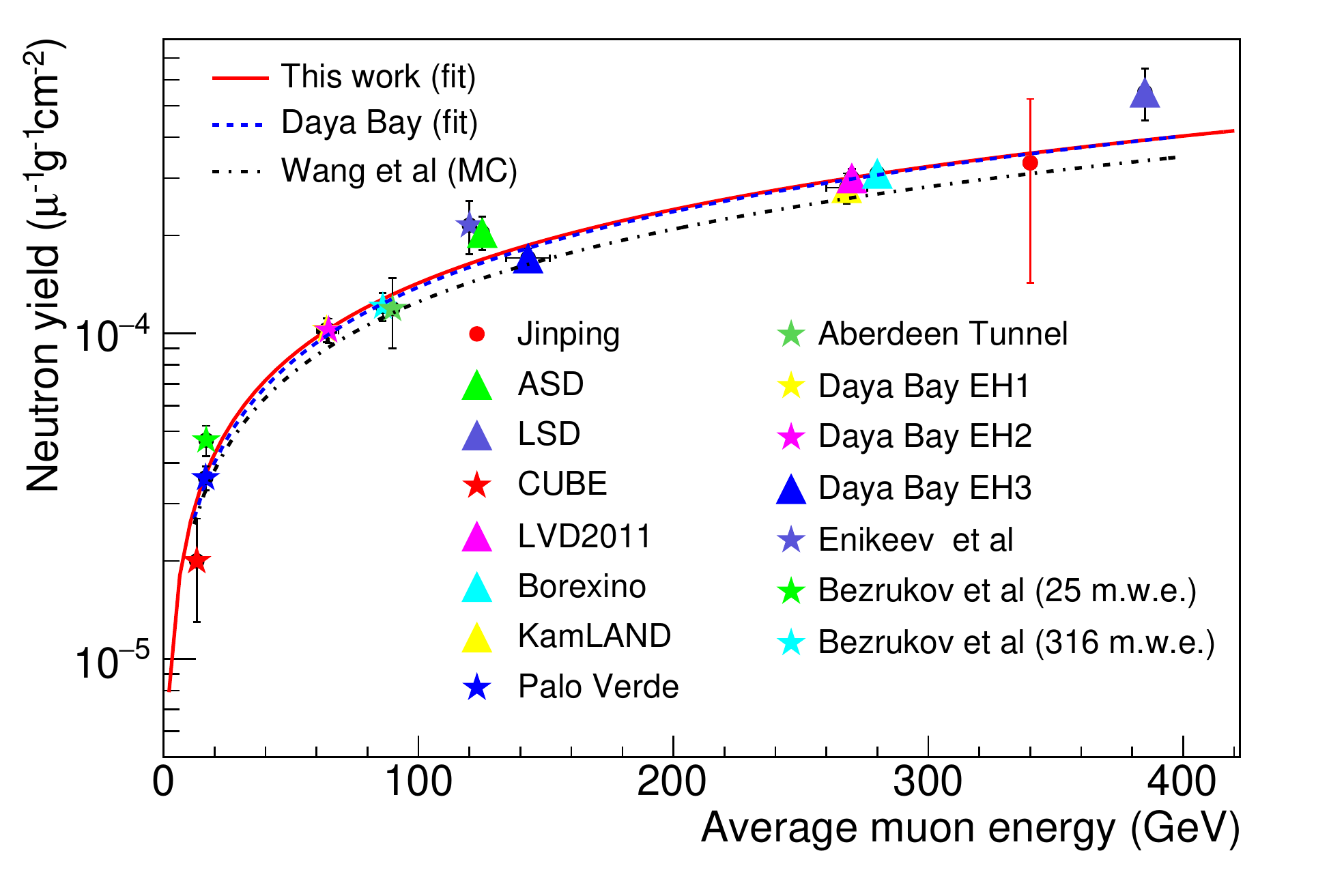}}
  \caption{(color online) Muon-induced neutron yield as a function of average muon energy from Jinping compared with other experiments. The solid red line shows the power-law fit to all the experiments, the blue dashed line is the fit result from Daya Bay~\cite{an2018cosmogenic}, and the black dashed line shows FLUKA-based predictions for the dependence of the neutron yield on muon energy from Ref.~\cite{wang2001predicting}.}
  \label{fig:LSYieldJP}
\end{figure*}

We compare the neutron yield measurements from different experiments in terms of the average muon energy, with various muon energies and angular distributions. Figure~\ref{fig:LSYieldJP} shows the measurement at CJPL and the results from CUBE~\cite{PhysRevC.52.3449}, Bezrukov et al.~\cite{bezrukov1973investigation}, Palo Verde~\cite{PhysRevD.62.092005}, Daya Bay EH1~\cite{an2018cosmogenic}, Daya Bay EH2~\cite{an2018cosmogenic}, Aberdeen Tunnel~\cite{PhysRevD.93.072005}, Enikeev et al.~\cite{enikeev1987hadrons}, ASD~\cite{enikeev1987hadrons,malgin2008neutrons,agafonova2013universal}, Daya Bay EH3~\cite{an2018cosmogenic}, KamLAND~\cite{abe2010production}, LVD2011~\cite{lvd-persiani2013measurement}, Borexino~\cite{Borexinobellini2013cosmogenic}, and LSD~\cite{aglietta1989neutron} as a function of average muon energy.

 Previous studies~\cite{wang2001predicting,kudryavtsev2003simulations,araujo2005muon,mei2006muon} have shown that the yield follows a power law of average muon energy
\begin{equation}
  Y_{n}=aE_{\mu}^{b}.
  \label{eq:powlaw}
\end{equation}
Figure~\ref{fig:LSYieldJP} shows the result of a global power-law fit to the experimental measurements solid red line, including the result from this study. The fit yields coefficients $ a=(4.52\pm0.55)\times 10^{-6}\mu^{-1}\rm g^{-1}cm^{2} $ and $ b= (0.75\pm 0.02)$.  Compared with the power-law fit from Daya Bay, this fit was consistent within the permitted error range, but it did not contribute significantly to the global fit owing to a large uncertainty.

We noted that the MC simulation~\cite{wang2001predicting} underestimates the data from LSD~\cite{aglietta1989neutron} at a high energy.  Mei and Hime~\cite{mei2006muon} attributed the deviation to the lack of experimental data for high-energy muon interactions with nuclei. However, the cross-section of muons calculated using the Bezrukov-Buagaev model~\cite{bezrukov1981nucleon} and used in the simulation~\cite{wang2001predicting} was consistent with the measurement by MACRO~\cite{kluck2015review} and ATLAS~\cite{alexa2003measurement}.  We expect the next hundred-ton JNE detector under construction to increase the exposure by at least 100 times, which will measure the neutron yield precisely to provide insights to the tension.

\section{Summary}

We studied the cosmic-ray data collected by the 1-ton prototype detector of JNE at CJPL-I. By analyzing the neutrons associated with the muons, we measured the neutron yield $ (3.44 \pm 1.86_{\rm stat.}\pm 0.76_{\rm syst.})\times 10^{-4}\mu ^{-1}\rm g^{-1}cm^{2} $ at an average muon energy of 340 GeV, corresponding to an upper limit of $ Y_{n}<7.78\times 10^{-4}\mu ^{-1}\rm g^{-1}cm^{2} $ at 99\% C.L.  A power-law fit of the neutron yield against average muon energy shows that our result is consistent with those of previous studies but lacks statistics to contribute significantly.  With the next hundred-ton JNE detector under construction, we foresee significantly a much more precise muon-induced neutron study in the near future.

\section{Acknowledgements}
We thank the anonymous referees for their critical comments and insightful suggestions.  We acknowledge Orrin Science Technology, Jingyifan Co., Ltd, and Donchamp Acrylic Co., Ltd, for their efforts
in the engineering design and fabrication of the stainless steel and acrylic vessels. Many thanks to the
CJPL administration and the Yalong River Hydropower Development Co., Ltd. for logistics and support.

\newpage
\bibliographystyle{unsrt}
\bibliography{bibfile}

\end{document}